\documentclass[twocolumn,showpacs,preprintnumbers,amsmath,amssymb]{revtex4}

\usepackage{graphicx}
\usepackage{dcolumn}
\usepackage{bm}

\def \ee{\end{equation}}
\def \be{\begin{equation}}

\preprint{arXiv:0707.4644}

\begin{document}

\title{Renormalization Group and Black Hole Production
in Large Extra Dimensions}

\keywords      {Large Extra Dimensions, Black Holes, Fixed Point}
\author{Ben Koch}
 \affiliation{
 Institut f\"{u}r Theoretische Physik, Johann Wolfgang Goethe -
Universit\"{a}t,\\
D--60438 Frankfurt am Main, Germany\\
}
\altaffiliation[Also at]{
FIAS--Frankfurt Institute for Advanced Studies, \\
Max von Laue-Str. 1,D--60438 Frankfurt am Main, Germany\\
}

\date{\today}

\begin{abstract}
It has been suggested
that the existence of a non-Gaussian
fixed point in general relativity might cure the ultraviolet problems
of this theory.
Such a fixed point is connected to an effective running
of the gravitational coupling. We calculate 
the effect of the running gravitational coupling on
the black hole production cross section
in models with large extra dimensions.
\end{abstract}

\pacs{04.60.-m}
\maketitle


The overwhelming success of the standard model
of particle physics in providing a consistent description of strong
and electroweak interactions encourages further steps
towards the unification of all forces. A next step
could be the unified description of the standard model and the general relativity,
which can be derived from the Einstein-Hilbert action.
In order to unify the two theories one needs
to solve two major problems.

One problem is encountered, when the assumption is made 
that general relativity and the standard model 
have their origins
in a single unified field theory X with a single unified mass scale
$M_X$. It is not understood why
the mass scales of general relativity ($m_{Pl}$) and of the standard model ($m_H$)
are so different in nature. In fact, a simple estimate shows that
\be\label{hier2}
\frac{m_H}{m_{Pl}}\sim 10^{-17}\;\quad.\;
\ee
This difference is known as the hierarchy problem.
Since the gravitational coupling is $G_N\sim 1/M_{Pl}^2$,
this leads to the question: " Why is gravity so weak as compared to
the other forces in nature?".
The hierarchy problem, which is shown in Eq. 
(\ref{hier2}), can be resolved by either
a Higgs mass of the order of $10^{19}$~GeV rather
than the expected few hundred GeV \cite{Kuti:1987nr} or
by lowering the Planck mass down to the TeV region.
However, a higher Higgs mass would aggravate 
the hierarchies between $m_H$ and the light fermions. This attempt
would then create a new hierarchy by eliminating the other. 
Lowering the Planck scale, however, would be 
much more desirable.
In the context of extra dimensions
there exist scenarios that
give rise to a lower Planck
scale and explain the difference in Eq. (\ref{hier2}).
Although, an explanation for the mass hierarchy does not imply 
that the unified theory X
has been found, it might give a useful hint 
on how to proceed.
In reference  \cite{ArkaniHamed:1998rs,Antoniadis:1998ig,ArkaniHamed:1998nn}  Arkani-Hamed, Dimopoulos, and Dvali
do this by assuming that the additional spatial dimensions are compactified
on a small radius $R$ and further demanding that all known particles live
on a $3+1$ dimensional sub-manifold ($3-$brane).
They find that the fundamental mass $M_f$ and the Planck mass $m_{Pl}$ 
are related by
\begin{eqnarray}
m_{Pl}^2 = M_f^{d+2} R^d \quad. \label{Master}
\end{eqnarray}
Within this approach it is possible to have a fundamental gravitational
scale of  $M_f\sim 1$~TeV.
The huge hierarchy between $m_H$ and $m_Pl$ would then come
as a result of our ignorance regarding extra spatial dimensions.

Another problem is the bad ultraviolet behavior of gravity.
The standard approach to the quantization of general relativity
with the metric $g_{MN}(x)$
includes perturbations $h_{M N}(x)$ (gravitons)
around the flat Minkowski metric $\eta_{M N}$ as the local quantum
degrees of freedom \cite{Weinberg:1972}
\be
g_{MN}(x)=\eta_{M N}+h_{M N}(x)\quad.
\ee
The standard loop expansion in the gravitational coupling,
fails because every new order in the perturbative
expansion brings new ultraviolet (UV) divergent
contributions to any physical process.
Since the Einstein-Hilbert action contains operators of mass dimension
higher than four, those divergences cannot be cured
by the standard renormalization procedure
used in the standard model.
A solution to this problem would be found
if one could show that the poor UV behavior is not
present in the full theory but only comes about due to the
expansion in the gravitational coupling.
According to this point of view, the full quantum gravity  would have
an energy dependent
asymptotically safe coupling constant with a non-Gaussian 
fixed point.
Such renormalization group
(RG) techniques have been successfully used for
a number of different problems 
\cite{Reuter:1992uk,Reuter:1996cp,Litim:2000ci,Litim:2001up,Litim:2003vp}.

A combination of these two approaches would
solve both, the hierarchy problem and the UV problem.
Since the existence of a non-Gaussian fixed point
in higher dimensional gravity was shown in \cite{Fischer:2006fz,Fischer:2006at},
an implementation of the running coupling into theories
with large extra dimensions 
\cite{Giudice:1998ck,Han:1998sg} was possible.
Moreover, RG effects on graviton production, graviton exchange, and
Drell-Yan processes in the context of extra dimensions
have been considered in \cite{Hewett:2007st,Litim:2007iu}.

In this paper we study how RG affects 
the possible production of microscopical black holes (BH),
which is probably
the most prominent collider signal for large extra dimensions.\\


A complete understanding of all BH properties is only possible
in a unified theory of quantum-gravity.
In the framework of large extra dimensions
the metric of a black hole with mass $M$ is given by
\begin{eqnarray}
\label{bhmetric}
ds^2&=&-\sqrt{1-\frac{16 \pi M}{(d+2) A_{d+2}m_{Pl}^2}r^{d+1}} dt^2\\ \nonumber
&&+
\frac{1}{\sqrt{1-\frac{16 \pi M}{(d+2) A_{d+2}m_{Pl}^2}r^{d+1}}}dr^2
+r^2 d\Omega_{d+2}^2\quad,
\end{eqnarray}
where $A_{d+2}$ is the $d+2$ dimensional sphere
\be
A_{d+2}=\frac{2 \pi^{\frac{3+d}{2}}}{\Gamma(\frac{3+d}{2})}\quad.
\ee
Due to the low fundamental scale $M_f\sim$~TeV and the hoop conjecture \cite{hoop},
it might be possible to produce such objects with mass of approximately
1 TeV in future colliders \cite{Banks:1999gd,Giddings:2000ay,Giddings:2001bu,Dimopoulos:2001hw}. 
This can only be the case when the invariant scattering energy $\sqrt{s}$
reaches the relevant energy scale $M_f$.
The higher dimensional Schwarzschild 
radius \cite{Giddings:2001bu,my} of these black holes is given by
\begin{equation} \label{ssradD}
R_H^{d+1}=
\frac{16 \pi (2 \pi)^d}{(d+2) A_{d+2}}\left(\frac{1}{M_{f}}\right)^{d+1} \; \frac{M}{M_{f}}
\quad .
\end{equation}
This would open up a unique possibility of studying quantum
gravity in the laboratory.
A semi-classical approximation for the 
BH production cross section is given by
\begin{eqnarray} \label{cross}
\sigma(M)\approx \pi R_H^2 \theta(\sqrt{s}-M_f)\quad,
\end{eqnarray}
where the theta function ensures that black holes are only produced
above the $M_f$ threshold.
This threshold condition is necessary because a black hole with $M<M_f$ would
not be well defined, as it would have 
for example a temperature $T>M$.
Such a cut will have
crucial significance in the RG approach to BHs.
The validity of this approximation has been debated in
\cite{Voloshin:2001fe,Voloshin:2001vs,Giddings:2001ih,Rizzo:2001dk,Jevicki:2002fq,Eardley:2002re,Rychkov:2004sf,Rychkov:2004tw,Kang:2004yk,Rizzo:2006uz,Rizzo:2006zb}.
Still, improved calculations including the diffuseness  
of the scattering particles (as opposed to point particles)
and the angular momentum  of
the collision (as opposed to head on collisions) as well as string inspired
arguments only lead to modifications of the order of
one \cite{Yoshino:2002tx,Solodukhin:2002ui,Ida:2002ez,Horowitz:2002mw}.
However, there are arguments that
the formation of an event horizon can never be observed \cite{Vachaspati:2006ki,Vachaspati:2007hr}.\\


Non-perturbative renormalization
is performed in Euclidian spacetime and has been successfully applied
to a variety of field theories such as quantum chromodynamics 
\cite{Reuter:1997gx} and  gravity 
\cite{Reuter:1996cp,Codello:2007bd}. In the case of gravity, it
offers a possible solution for the problem of non renormalizable
UV divergences in the perturbative approach.
This solution appears
due to the possible existence of a Gaussian fixed point in the UV regime
and a non-Gaussian fixed point in the infrared regime.
 
The main idea in this approach is that
it is possible to introduce an infrared cutoff operator in
the theory, which leaves the effective Lagrangian 
invariant under general diffeomorphism transformations.
After a gauge fixing it was possible to derive
an exact evolution equation for the effective action \cite{Reuter:1992uk}.
Recently this method has been generalized to
more than three spatial dimensions \cite{Fischer:2006fz,Fischer:2006at} and
applied to models with large extra dimensions \cite{Hewett:2007st,Litim:2007iu}.
Both approaches show that cross sections in extra dimensional theories, which 
originally had a non-unitary behavior for $\sqrt{s}\gg M_f$,
are now well defined in this high energy limit.

In \cite{Litim:2003vp,Hewett:2007st} the running gravitational
coupling $\tilde{M}_f(\sqrt{s})$ is
\be\label{Mfhut}
\tilde{M}_f(\sqrt{s})=M_f \left(
1+\left( \frac{s}{t^2 M_f^2}
\right)^{\frac{d+2}{2}}
\right)^{\frac{1}{d+2}}\quad.
\ee
It
turns out that $t$, which can be
obtained from a series of non-trivial integrals, 
is of the order one in the relevant region
of parameter space \cite{Hewett:2007st}.
This running coupling has two asymptotic regimes.
For low energies $M_f\gg\sqrt{s}\gg 1/R$
one obtains  $\tilde{M}_f(\sqrt{s}\rightarrow 1/R)=M_f$,
whereas, for very high energies $\sqrt{s}\gg M_f$
one sees that the effective higher dimensional Planck mass
diverges $\tilde{M}_f(\sqrt{s}\gg M_f)\approx \sqrt{s}/y\approx \sqrt{s}$.
This diverging mass scale corresponds to a vanishing gravitational
coupling, which is exactly the desired asymptotic safety.\\


For three spatial dimensions the RG effects 
on the decay of astronomical black holes have already 
been discussed in \cite{Bonanno:2006eu}.
Surprisingly enough, it turned out that RG effects slow down 
the Hawking evaporation until a stable black hole remnant is formed. 
A prediction,
which was also made using different arguments for the extra dimensional
BH's at the large hadron collider \cite{Koch:2005ks,Hossenfelder:2005ku,Humanic:2006xg,Koch:2007um}.
However, before considering the decay of mini black holes,
the RG effects on the formation of black holes should be studied.

\begin{figure}[htb]
\includegraphics[width=8.5cm]{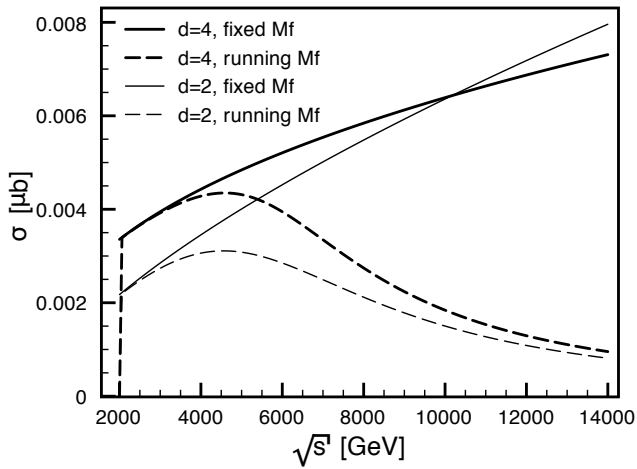}
\caption{
BH area in $\mu b$ for $M_f=2000$~GeV and $t=3$ as a function of $\sqrt{s}$. 
 \label{farea}}
\end{figure}

\begin{figure}[hbt]
\includegraphics[height=6.5cm]{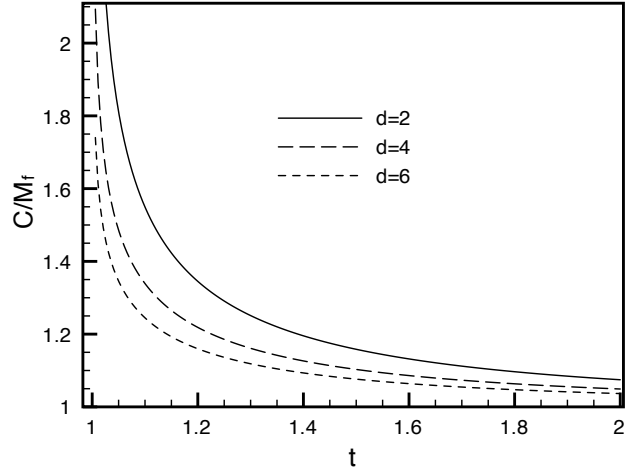}
\caption{
Threshold $C$ normalized to $M_f$ as a function of $t$ for $d=2,4,6$.
 \label{tmp}}
\end{figure}

\begin{figure}[htb]
\includegraphics[width=8.5cm]{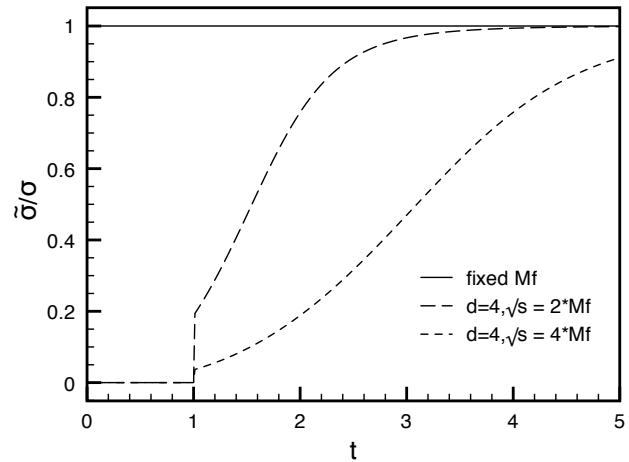}
\caption{
Dependence of the normalized cross section $\tilde{\sigma}/ \sigma$ on the
 regularization parameter t for $d=4$ and for $\sqrt{s}=2 M_f$ $(\sqrt{s}=4 M_f)$.
 \label{sigVont}}
\end{figure}

This can be done relatively straight forward,
by plugging Eq. (\ref{Mfhut}) into Eq. (\ref{cross}).
One finds the cross section for
the case of a running gravitational coupling
\begin{eqnarray}\label{crosshut}
\tilde{\sigma}(\sqrt{s}) \approx&
\frac{\pi}{ \tilde{M}^2_f(\sqrt{s})}&
\left(\frac{16 \pi(2 \pi)^d \sqrt{s}}{(d+2) A_{d+2}
\tilde{M}_f(\sqrt{s})}\right)^{2/(d+1)}\\ \nonumber
&\cdot &\theta(\sqrt{s}-\tilde{M}^2_f(\sqrt{s}))\quad.
\end{eqnarray}
The running Planck scale in Eq.
(\ref{crosshut}) has three consequences.
First, for $\sqrt{s}\sim M_f$ the higher dimensional
Planck mass is  enhanced by a factor of $(1+t^{-d-2})^{1/(d+2)}$
which lowers the area only by a factor of 
$(1+t^{-d-2})^{-2 (d+2)^2/(d+2)}$.
As long as $t$ is of order one, this corresponds to a change of
just a few percent.
Secondly, for very high energies $\sqrt{s}\gg M_f$,
the asymptotic safety wins over the increased energy
and the BH area goes to zero $\sigma \sim 1/s\; (\sqrt{s}/M_f)^{-1/(d+1)}$.
In Fig. (\ref{farea}) the BH area is shown
as a function of $\sqrt{s}$ for the cases with and without RG effects.
The area drops off when $\sqrt{s}$ is large
and, therefore, looks like a ``black hole resonance".
%
The third consequence of the running $\tilde{M}_f$ is on
the threshold condition. The theta function
in (\ref{crosshut}) gives
\be
\sqrt{s}=\left( \frac{t^2}{t^2-1}\right)^{1/(d+2)}M_f\quad,
\ee
which shows that there are dramatic consequences for the standard 
picture of BH production 
as soon as $t$ is of order one.
For large $t\gg1$ the threshold $C$ will basically be just slightly raised above
$M_f$ but as soon as $t$ approaches one the shift increases
until when $t\leq 1$ black holes no longer produced.
The sharp behavior of the threshold $C$ as a function of $t$
is shown in Fig. (\ref{tmp}) for various $d$.
This cutoff behavior means that the cross section (\ref{crosshut})
approaches the standard cross section (\ref{cross}) only in 
the limit where $t\rightarrow \infty$
and is zero for $t\leq1$,
as shown in Fig. $\ref{sigVont}$.\\


We applied RG techniques to the black hole production scenario
in the context of large extra dimensions.
We found two surprising effects.
First, the area of the black hole,
which is of the same order of magnitude as the production cross section,
is not only UV safe (as it was observed
in standard scattering cross section) but it is damped
so strongly that it goes to zero.
Secondly, the truncation parameter $t$, which does not play
an important role in the qualitative standard scattering cross section
picture, is very  important for the BH threshold.
Moreover, BH production could be completely 
forbidden for $t\leq 1$,
 which according to \cite{Hewett:2007st} is perfectly possible.

In the simplest picture
of BH production the RG has dramatic consequences. 
Further study is needed to check whether the results
obtained here remain valid after an
improved formulation of the BH threshold, the RG solutions, and
the truncation parameter $t$.
If the results do not change in a more detailed formulation
and $t$ can be determined to be smaller than one
no black holes will be produced at
future colliders regardless if large extra dimensions exist or not.\\ \\
The author wants to thank Jorge and Jacki Noronha, 
Christoph Rahmede, Sabine Hossenfelder, and Daniel Litim
for their remarks and interesting discussions.\\
 This work was supported by GSI.\\

\end{document}